\documentstyle[12pt,epsfig]{article} 
\newcommand{\ba}{\begin{array}}
\newcommand{\ea}{\end{array}}
\newcommand{\bd}{\begin{displaymath}}
\newcommand{\ed}{\end{displaymath}}
\newcommand{\be}{\begin{equation}}
\newcommand{\ee}{\end{equation}}
\newcommand{\bea}{\begin{eqnarray}}
\newcommand{\eea}{\end{eqnarray}}

\def\barr{\begin{array}}
\def\earr{\end{array}}

\newcommand{\beq}{\begin{eqnarray}}
\newcommand{\eqn}{\end{eqnarray}}

\def\l{\lambda}

\def\etal{ {\em et al.}~}

\def\q2 {q^2}

\def\N10{\widetilde \chi_1^0}
\def\Cp1{\widetilde \chi_1^+}
\def\Cm1{\widetilde \chi_1^-}
\def\C1pm{\widetilde \chi_1^\pm}

\begin{document}
\begin{titlepage}

%\begin{flushleft}
%\today\\
%version-5
%\end{flushleft}
\begin{flushright}
{\large OITS-726}
\end{flushright}

\begin{center}
{\Large\bf Invisible decays of Higgs and other 
mesons in models with singlet neutrinos in large extra dimensions}\\[10mm]
N.~G.~Deshpande{\footnote{email: desh@oregon.uoregon.edu}}, 
Dilip Kumar Ghosh{\footnote{email: dghosh@physics.uoregon.edu}} 
\\[4mm]
{\em Institute of Theoretical Science \\
     5203 University of Oregon\\
Eugene OR 97403-5203} \\[10mm]
\end{center}
\begin{abstract}
In light of current atmospheric neutrino oscillation data, we revisit the 
invisible decay of the standard model Higgs boson and other pseudoscalar 
mesons which can be enhanced because of large number of KK modes 
in models with right handed singlet neutrinos in large extra 
dimensions. We find that the invisible decay rate
of Higgs can be as large as $H\to b \bar b $ decay rate only for a 
very restricted region of parameter space. This parameter space is even further
restricted if one demands that the dimensionless 
neutrino Yukawa coupling $\l$ is $O(1)$. 
We have also studied the scenarios where singlet neutrino propagate in a 
sub-space, which lowers the string scale $M_{\ast}$ and keeps neutrino 
Yukawa coupling $O(1)$. We have also considered decays of other spin-$0$
mesons to $\nu \bar\nu$ and found the rates to be too small for 
measurement.  

\end{abstract}
 
\end{titlepage}

\vskip 1 true cm

\newpage

\renewcommand{\thepage}{\arabic{page}}
\setcounter{page}{1}

The concept of large extra dimensions and TeV scale quantum 
gravity~\cite{ADD} introduced by ADD has attracted a lot of 
attentions recently. In this 
scenario, one has $\delta $ additional spatial dimensions of size $R$ in 
which gravity propagates 
whereas the standard model particles with chiral fermionic content
are confined to the usual $4$-dimensions $(4D)$. The effective $4D$ Planck 
scale, $M_{Pl}\sim 2.4\times 10^{18}$~GeV , is related to the {\it 
fundamental } Planck scale $M_{*}$ in $(4+\delta)$ dimension by
\beq
M^2_{Pl}\sim R^{\delta } M^{\delta +2}_{*}
\eqn 
Thus, for the large extra-dimensions, it is possible to have a 
{\it fundamental} scale $M_{*}$ as low as a TeV~\cite{ADD}, helping to 
resolve the
gauge hierarchy problem of the standard model.  
The size of the extra dimensions can be as large as $\sim$mm for $\delta =2$.
Another interesting aspect of this scenario is the generation of a small 
neutrino mass~\cite{neut1,neut2}. The relatively small value of $M_{*}$ 
indicates the possibility of a seesaw mechanism, with right-handed (RH) 
singlet neutrinos also propagating in the full $4+\delta$
dimensions with gravity~\cite{neut3,neut4,neut5,neut6,neut7,neut8}. For illustration let us 
assume the case of a single extra-dimension labeled by $y$ 
($x$ labels the usual $4D$)~\cite{desh1}. 

A massless Dirac fermion $N$
which is a standard model singlet lives in $5D$. The $\Gamma$ 
matrices in $5D$ can be written as
\begin{equation}
\Gamma ^{\mu} = 
\left(
\begin{array}{cc}
0 & \sigma ^{\mu} \\
\bar{ \sigma } ^{\mu} & 0
\end{array}
\right), 
\Gamma ^5 = 
\left(
\begin{array}{cc}
i & 0 \\
0 & -i
\end{array}
\right).
\end{equation}
The Dirac spinor $N$ in the Weyl basis can be written as
\begin{equation}
N = 
\left(
\begin{array}{c}
\psi \\
\bar{\chi}
\end{array}
\right),
\end{equation}
where $\psi$ and $\chi$ are $2$-component complex spinors
(with mass dimension $2$).
The $5D$ kinetic term for $N$ is
\begin{equation}
S_{\hbox{free}} = \int d^4 x dy \; \bar{N} \left( \Gamma ^{\mu}
\partial _{\mu} + \Gamma ^5 \partial _y \right) N.
\label{kinetic}
\end{equation}
and the interaction action is 
\begin{equation}
S_{\hbox{int}} = \int d^4x \; \frac{\l}
{\sqrt{M_{\ast }}} \; l (x) H^{\ast} (x)
\psi (x, y = 0),
\label{Lint}
\end{equation}
where, $\ell = ( e, \nu ) $ is the standard model lepton doublet and 
$H$ is the Higgs doublet and $\l $ is a dimensionless Yukawa 
coupling in $5D$. We have assigned $N$ and $\ell$ opposite lepton number,
so that the lepton number is conserved in Eq.(5).
 
In the effective $4D$ theory, $N$ appears as a tower of Kaluza-Klein 
(KK) states:
\begin{equation}
\psi = \sum _n \frac{1}{\sqrt{R}} \psi ^{(n)} (x)
e^{iny / R},
\label{kk}
\end{equation}
where $\psi ^{(n)}$ are $4D$ states.   
Similarly, $\chi$ has a KK tower, $\chi ^{(n)}$.
Thus, in $4D$, we have the effective action: 
\beq
S_{\hbox{free}} &=& \int d^4x \sum _n \left[ \bar{\psi}
^{(n)} \bar{\sigma}^{\mu}
\psi
^{(n)} + \bar{\chi}
^{(n)} \bar{\sigma}^{\mu}
\chi
^{(n)} + \left( \frac{n}{R} \psi ^{(n)} \chi ^{(n)} + \hbox{h.c.} 
\right) \right],\\
S_{\hbox{int}} &=& \int d^4x \; \sum _n \; \l
\frac{M_{\ast}}{M_{Pl}} l (x) H^{\ast} (x) \psi ^{(n)} (x).
\eqn              
where $n / R$ is the Dirac mass for the KK states.

Thus, we see that the $4D$ neutrino Yukawa coupling is suppressed
by the volume of the extra dimensions, or in other words, by the 
ratio $M_{\ast}/M_{Pl}$. When $H$ gets a VEV ( $ v \approx 246 $ GeV ), we get
Dirac mass for standard model neutrino $(\nu)$ denoted by $m$:
\begin{equation}
m \approx \frac{\l}{\sqrt{2}}\frac{M_{\ast}}{M_{Pl}} v~ 
\sim~\l\frac{M_{\ast}}{{\rm TeV}} 10^{-4}~{\rm eV}.
\label{m-lam}
\end{equation}
However, this standard model neutrino (which is dominantly the lightest 
neutrino) has a small mixture ($\sim mR/n$) of heavier neutrinos. With this,
the lightest neutrino mass is modified to ( we extend the formula to $\delta$
extra-dimensions)
\begin{equation}
m_{\nu}\approx \frac {m}{\sqrt{1+{\frac{m^2}{M^2_{*}}{\frac{M^2_{Pl}}{M^2_{*}}}
{\frac{2\pi^{\delta/2}}{\Gamma(\delta/2)(\delta -2)}}}}  }
\label{m-mnu}
\end{equation}  
Now, for a given value of $M_{*}$ and $\delta$ the upper limit of 
$m_{\nu}$ is $m^{max}_{\nu}\approx \frac{M^2_{*}}{M_{Pl}}
{\sqrt{\frac{\Gamma(\delta/2)(\delta-2)}{2\pi^{\delta/2}}}} $, 
( for $\delta = 2, \delta-2$ is replaced by $1/ln(M_{Pl}/M_{*})$). 
We then get 
\beq
m_{\nu}\approx \frac{m}{   \sqrt{1 + (\frac{m}{m^{max}_{\nu}})^2}        }
\eqn
%\section{Invisible decay of Higgs}
\vskip .1in      

{\large \bf Invisible decay of Higgs}

It has been shown in Ref.\cite{wells} that the standard model Higgs can decay 
into $\nu_L \bar \nu_R^i$ with the strength 
$\sim \frac{\l}{\sqrt{2}}\frac{M_{\ast}}{M_{Pl}} $, which looks like rather 
a small number. However, one can a get large enhancement in the decay rate when 
one sums the KK excitations of right-handed neutrino states with masses 
below the Higgs mass
$(M_H)$. This summation is proportional to the volume 
$R^\delta$ of the extra-dimensional space, with a momentum-space factor of
order $(M_H)^\delta $. So, the rate is enhanced by the factor 
$(M_H R)^\delta $. 
Therefore the sum of partial widths of the Higgs into $KK$ excitations 
of neutrinos is of the order of :
\beq
\sum_{i} \Gamma (H \to \nu_L\bar\nu^i_R)\sim \frac{M_H}
{16\pi}Y_{\nu}^2(M_HR)^\delta
\eqn
where, $Y_{\nu}\sim {\frac{\lambda}{\sqrt{2}}}{\frac{M_{*}}{M_{Pl}}} $.
To quantify the invisible decay rate of the Higgs boson, following the
notation of Ref.\cite{wells}, we define the ratio 
$\kappa (\equiv \sum_{i} B(H \to \nu_L\bar\nu^i_R) / B(H\to b \bar b) )$, which
 can be expressed as :
\beq 
\kappa \simeq \frac{m^2}{3 m_b^2}\left ( \frac{M_H}{M_{\ast}} \right )^\delta 
\left ( \frac{M_{Pl}}{M_{\ast}}\right)^2 
\eqn
From this expression it is clear that $\kappa$ increases with $m$ and $M_H$,
while it decreases with increase in $\delta $ and $M_{\ast}$. Note that 
$\kappa$ depends on $m$ rather than $m_{\nu}$ directly.  

We now evaluate the ratio $\kappa$ in light of the latest atmospheric neutrino
oscillation data, which suggest 
$\Delta m^2_{atm}= {( 1.5-4.0 ) \times 10^{-3}~{\rm eV}^2 } $
\cite {rjw}. 
%In our analysis we fix $m_{\nu}\sim \sqrt{\Delta m^2_{atm}}$
%and then solve form $m$ from Eq.(10) in terms of $M_{\ast}$ and $\delta $. 
%We will consider two cases, $(a)$ $m^2_{\nu}\sim 1.5\times 10^{-3}$,
%$(b)$ $m^2_{\nu}\sim 4.0\times 10^{-3}$. 
In our analysis we fix the highest neutrino mass which dominates 
$H\to \nu\bar\nu $ rate to be $m_\nu\sim \sqrt{\Delta m^2_{atm}} $ and then
solve for $m$ from Eq. (\ref{m-mnu}) in terms of $M_{\ast}$ and $\delta $. 
The seesaw mechanism in Eq. (\ref{m-lam}) clearly favors hierarchical 
neutrino masses rather than degenerate mass spectrum ( which will need 
fine tuning of $\l$). We consider two cases that respect the allowed range
of $\Delta m^2_{atm}$, $(a)$~$m^2_\nu = 1.5\times 10^{-3}~{\rm eV}^2 $
and $(b)$~$m^2_\nu = 4.0\times 10^{-3}~{\rm eV}^2 $. Another important 
consideration is the value of $\l$ that ensues for a given $m$ and $M_{\ast}$.
Clearly, the value of $\l$, which is the higher dimensional Yukawa coupling, 
should not exceed $\sim 10-20 $, otherwise we will be in a non perturbative
regime. This is a serious constraint on allowed value of $M_{\ast}$ for a
given $\delta$. 

In Figure 1 $(a)$ we show the variation of $\kappa $ with $M_{\ast}$ for three 
choices of extra dimensions $\delta =3, 4~{\rm and }~5 $ keeping 
$m^2_{\nu} = 1.5\times 10^{-3} $. 
The solid, broken and dotted curves correspond to 100, 150 and 200 GeV 
Higgs mass respectively. Left end of each curve has been truncated by using 
a constraint $m \leq 4 m_\nu $. Value of $m$ higher than this correspond to 
unacceptably large $\l $. From this Figure, it is clear that for $\delta=4$
and $5$ the invisible width is only a tiny fraction of $H\to b\bar b$ rate. 
For $\delta = 3 $, the value of $\kappa$ is at most $1,3 $ or $7$ for the 
Higgs masses 100, 150 and 200 GeV respectively, and less than $1\%$ for most 
of the parameter space. 
 
Similar variation is shown in Figure 1 $(b)$, however, for 
$m^2_{\nu}\sim 4.0 \times 10^{-3}$. The situation is better in this case, 
since, higher value of $m_{\nu}$ lead to larger value of $m$
in the ratio $\kappa$. In this case also we have truncated each curve 
on the left by imposing the same condition as in Figure $1(a)$. 

Before we proceed, we would like mention two points here. 
Firstly, beyond $M_H = 150$ GeV, the dominant mode of Higgs decay is 
$H\to W W^\ast $, thus the $H\to b\bar b$ branching ratio is very small, 
and any $H \to \nu \bar \nu$ branching ratio will also become small.  
However, for the purpose of comparison we have presented 
the case of 200 GeV Higgs mass.
One should also note that for the value of $M_{\ast}$ at which 
$\kappa $ is $O(1)$ and larger, corresponding Yukawa coupling $\l $ is 
around 80 for $\delta = 3.$ Such a large value of $\l$ may not be accepted 
from the perturbative point of view. 

For a given value of $m_\nu$ and $M_{\ast}$, the value of $\l$ can be 
obtained from Eq. (\ref{m-lam}) and Eq. (\ref{m-mnu}):
\beq
\l = \frac{\left( m_\nu/M_{\ast}\right) 10^{16}}
{\sqrt{1 -\frac{m^2_\nu}{M^2_{\ast}}\frac{M^2_{Pl}}{M^2_{\ast}}\frac{2\pi^{\delta/2}}{\Gamma(\delta/2) \left(\delta -2 \right )}}}
\eqn
It is easy to show that if we limit $\l\leq 10 $ (to be in perturbative regime)
, for $m^2_\nu= 1.5\times 10^{-3} {\rm eV}^2 $ we are restricted to 
$M_{\ast}\geq 40 $ TeV and for $m^2_\nu= 4.0\times 10^{-3} {\rm eV}^2 $,
$M_{\ast}\geq 64 $ TeV. The dependence on $\delta$ is very week on these 
bounds. Thus for such high values of $M_{\ast}$ the ratio 
$\kappa$ is highly suppressed.

For completeness we discuss the case of degenerate neutrinos, separated by 
$\Delta m^2 $ consistent with the atmospheric and solar neutrino data. 
Recently WMAP \cite{wmap} provided important new information on cosmic 
microwave background anisotropies. After combining the data from 2dF Galaxy
Redshift Survey, CBI, and ACBAR\cite{acbar}, WMAP places stringent limits on 
the contributions of  neutrinos to the energy density of the universe
~\cite{gbhat} 
\beq
\Omega_{\nu} h^2  = \frac{\sum_{i} m_{\nu_i}}
{93.5~{\rm eV}}< 0.0076~~~
(95\%~{\rm C.L.}) , 
\eqn
which implies 
\beq 
\sum _i m_{\nu_i} < 0.71  eV
\eqn
for
a single active neutrino, or $m_\nu < 0.23 $ eV for three degenerate 
neutrinos. Using the value $m_\nu = 0.23 $ eV and multiplying Eq.~(13) 
by $3$ for 
three families we can calculate $\kappa$. This is shown in Figure $2$ .
The general behavior of $\kappa $ as a function of $M_{\ast}$ is similar
to Figure 1. The only point to be noted here is that, because of 
heavier neutrino mass, the allowed range of $M_{\ast}$ is much higher 
than the previous case. If we now impose the condition that $\l \leq 10 $
to satisfy the perturbative condition, $M_{\ast}$ becomes 
too heavy $\sim 200 $ TeV. For such a large value of $M_{\ast}$,   
$\kappa$ drops below $10^{-4}$ for $M_H = 100 $ GeV and $\delta = 3$,
, and is essentially unmeasurable. The situation is worse for 
higher values of $\delta $.  

In summary, we see from Figures $1(a), (b)$ and Figure $2$ that  
$\kappa $ of the order $O(1)$ can arise only when  $M_{\ast}\sim 20-30 $ TeV. 
However, as noted, this implies $\l > 70 $, making the theory non 
perturbative. There are two ways by which one can evade this problem.
Either one should take $\l\sim O(1)$, in that case, $ M_{\ast}$ will be
$\geq 100 $ TeV for $m_{\nu}\sim \sqrt{\Delta m^2_{atm}}$ and 
$M_{\ast} \geq 200 $ TeV for $m_{\nu} = 0.23 $ eV. This is perfectly fine,
but for such a large value of $M_{\ast}$, the invisible decay width of Higgs 
will be negligibly small compared to $H\to b \bar b $. The second choice 
suggested by ~\cite{neut2,Agashe-Wu} is to consider that the 
singlet neutrino propagates in a sub-space $(\delta_\nu )$ 
of the full extra-dimension $(\delta )$ 
where gravity propagates. Assuming that all extra 
dimensions are of the same size $R$, in this case, the Dirac mass for the
standard model neutrino now becomes
\beq
m \sim \l v \left (\frac{M_{\ast}}{M_{Pl}}\right )^{\frac{\delta_\nu}{\delta}}
\eqn
Thus for $\delta_\nu=5$ and $\delta = 6$ and with $M_{\ast}\sim {\rm TeV}$,
$\l \sim O(1)$, we obtain $m^2 \sim \Delta m^2_{atm}$. It has been shown
by Agashe and Wu~\cite{Agashe-Wu} that for the above choices of 
$\delta_{\nu}$ and $\delta$, the
constraints on $M_{\ast}$ from $BR(\mu \to e\gamma)$ and $\pi \to e\bar\nu, 
\mu \bar \nu$ decays can be significantly weakened as compared to the 
minimal model, allowing the scale $M_{\ast}\sim {\rm TeV}$.  
In this case, the maximum value of the physical neutrino mass for a given
$\delta_{\nu}, \delta $ and $M_{\ast}$ is given by \cite{Agashe-Wu}:
\beq
m^{max}_{\nu}\approx M_{\ast}\left(\frac{M_{\ast}}
{M_{Pl}}\right )^{\frac{\delta_\nu}{\delta}}
\sqrt{\frac{\Gamma\left (\delta_\nu/2\right)\left(\delta_\nu-2\right) }
{2\pi^{\delta_\nu/2}}}
\label{sub_dim}
\eqn
Following Eq.~(\ref{sub_dim}), one can invert Eq. (11) to get the neutrino 
mass parameter $m$ which enters in the Higgs decay width. 
In sub-space, by replacing $\delta \to \delta_\nu$ and 
$ (M_{Pl}/M_{\ast})^2 \to (M_{Pl}/M_{\ast})^{2(\delta_\nu/\delta)}$ one can 
get the ratio $\kappa$ in terms of the neutrino mass parameter $m, 
\delta_\nu, \delta, $ and $ M_{\ast}$. 
\beq
\kappa \simeq\frac{m^2}{3 m_b^2}
\left (\frac{M_H}{M_{\ast}}\right )^{\delta_\nu}
\left ( \frac{M_{Pl}}{M_{\ast}} \right )^{2 \left(\delta_\nu/\delta\right )}
\eqn
We now compute the ratio $\kappa$ as a function of $M_{\ast}$ keeping 
$\delta_\nu= 5 $ and $\delta=6$. 
In Figure 3 $(a)$ and $(b)$ , 
we show $\kappa$ as a function of $M_{\ast}$ for 
$m^2_{\nu}= 1.5 \times 10^{-3}~{\rm eV}^2 $ and $4.0\times 10^{-3}~{\rm eV^2}$
respectively. The choice of the Higgs masses is the same as before. 
Comparing Figure $(3)$ with the earlier Figures, it is clear that 
$M_{\ast}$ can now be as low as $1$ TeV. This value also is allowed by 
other experimental constraints~\cite{Agashe-Wu}. 
From Figure  $3~(a)$, one can see that $\kappa \sim 1$ only for 
$M_H = 200 $ GeV and at $M_{\ast} = 1 $ TeV. 
As $M_{\ast}$ increases $\kappa$ decreases significantly.
While in Figure  $3~(b)$, which 
corresponds to the higher value of the physical neutrino
mass, $ m^2_{\nu}= 4.0 \times 10^{-3} {\rm eV}^2 $, 
$\kappa$ can be $\sim 1 $ even for $M_H = 150 $ GeV. But for $M_H = 100 $ GeV 
it is always less than $1$. For $M_H = 200 $ GeV, $\kappa$ can be 
as large as $\sim 10 $, though for very small window in 
$M_{\ast} \sim 1- 1.3 $ TeV. Similar variation of $\kappa $ with $M_{\ast}$ 
is shown in Figure $4$ in the degenerate neutrino mass scenario. In this case,
we find that $\kappa $ can be larger than $1$ for $M_H = 200 $ GeV. 
However, as mentioned before, this value of Higgs mass will not serve 
the purpose of looking for invisible decay modes of the Higgs boson. 
Hence, for
practical purpose, we should look at values of $\kappa $ for $M_H $ up to
150 GeV. It turns out that for $M_H = 150 $ GeV, $\kappa \sim 0.8 $ for 
$M{\ast}\sim 2.5 $ TeV, whereas, for $M_H = 100 $ GeV, $\kappa $ can be
at most $0.1 $. 

In these two analysis, we have shown that only in a very small region of 
parameter space can the invisible decay of Higgs be as large as 
$H\to b \bar b $ decay mode. The main restriction comes from the 
perturbative constraint on the Yukawa coupling $\l$. 

For completeness we also discuss the case of asymmetric dimensions
~\cite{neut4, neut7}.
In this scenario neutrinos propagate in a sub-dimensional
space of dimension $\delta_\nu$ of size $R$, whereas gravity propagates in 
space of dimension $\delta $. The extra 
$(\delta-\delta_\nu)$ has a size $r$ with $(r << R)$. In such a scenario
Eq. (1) becomes 
\beq
M^2_{Pl}\sim R^{\delta_\nu } M^{\delta +2}_{*} r^{(\delta - \delta_\nu)}
\eqn
In this case, the Dirac mass for the standard model neutrino becomes 
\beq
m \approx \frac{\lambda v }{\sqrt{2\left(M_{\ast} R\right)^{\delta_\nu}}}
\eqn
To satisfy the constraints from supernova $ 1987a$, the scale 
$1/R > 10 $ KeV~\cite{neut6}. For such a value of $1/R$ 
the mixing angle $(\sim mR/n)$ of any KK state with the standard
model neutrinos becomes negligibly small and $m \approx m_{\nu}$. 

One can obtain $m_{\nu}^2 \sim \Delta m^2_{\rm atm}$, for 
$\delta_{\nu} = 3$, $\lambda \sim O(1) $, $M_{\ast}\approx 4 $TeV 
( for $R^{-1}\approx 10 $ KeV) and $ M_{\ast}\approx 8 $ TeV 
( for $R^{-1}\approx 25 $ TeV). For $\delta_\nu > 3 $, the neutrino mass
$m_\nu$ is highly suppressed as seen from above expression for $m$~(Eq. 21).
   
In this scenario, the ratio $\kappa $ as defined in Eq.(13) turned out to be:
\beq
\kappa &\simeq & \left ( \frac{m^2}{3 m_b^2} \right ) 
\left( M_H R \right )^{\delta_\nu}\\
       &\simeq & \left (\frac{\lambda^2 v^2}{6 m_b^2} \right) 
\left( \frac{M_H}{M_{\ast}}\right )^{\delta_\nu}
\eqn
To study the invisible Higgs decay in this scenario, we fix $\lambda =1 $
and take the same values of Higgs mass as before.
We find $\kappa = 0.17, 0.6 $ and 1.3  for $M_H = 100, 150 $ and 200 GeV
respectively for $M_{\ast} = 1.5$  TeV. 
From the Eqs.(22) and (23) it is clear that $\kappa $ 
decreases as $\delta_\nu$, and /or $M_{\ast}$ increases, while it increases
as the higher dimensional Yukawa coupling $\lambda $ increases.

Now we will discuss the observability of this 
kind of invisible decay of some pseudoscalar mesons. 
  
\vskip .1in      

{\large \bf Neutral Pion decays}

We begin our analysis with the neutral pion. 
The effective Lagrangian for the process $\pi^0 \to \nu_L \bar\nu^i_R$ is
in the standard model through $Z$ exchange is :
\beq
{\cal L}_{\rm eff} = \frac{i G_F}{\sqrt{2}} F_{\pi}m\bar\nu\gamma_5\nu
 \Phi_{\pi}
\eqn

where, $G_F$ and $F_\pi$ are the Fermi coupling constant and pion decay 
constant respectively. 

Using this effective Lagrangian we one determine the decay width of 
pion into $\nu_L \bar \nu^i_R$ in the minimal model:
\beq
\Gamma (\pi^0 \to \nu_L \bar\nu^i_R) = \frac{G_F^2F^2_{\pi}m^2 m_{\pi}}
{16\pi} \left (\frac{m_\pi}{M_{\ast}}\right )^\delta \left( 
\frac{M_{Pl}}{M_{\ast}}\right )^2
\eqn
We have computed the branching ratio $BR(\pi^0 \to \nu_L \bar \nu^i_R )$ 
for $ m^2_{\nu}= 4.0 \times 10^{-3} {\rm eV}^2 $ and found the result is
 $\sim 10^{-25}$ for $\delta = 3, M_{\ast}= 200 $ TeV and $\l \sim 11 $.
This predicted branching ratio is much smaller than the 
experimental upper limit $8.3 \times 10^{-7}$ at $90\%$ C.L.\cite{pdg}.

The branching ratio can be higher by $4-5$ order of magnitude for lower
values of $M_{\ast}\sim 45 $ TeV, but such a low value of $M_{\ast}$ 
correspond to unacceptably large $\l \sim 200 $. In the case of degenerate 
neutrinos, the above branching ratio does not change significantly, so 
we do not present any numerical results here.

Next, we compute the above decay widths in the sub-space 
scenario. In this scenario as shown earlier one can have $\l$ of order 
one and also $M_{\ast}$ of about few TeV. With the following replacement 
$\delta \to \delta_\nu$ and 
$(M_{Pl}/M_{\ast})^2 \to  (M_{Pl}/M_{\ast})^{2 (\delta_\nu/\delta )}$ one can
rewrite the decay width $\Gamma \left(\pi^0 \to \nu_L \bar\nu^i_R\right )$
in sub-space model.
In this case, we find that for $m^2_{\nu}= 4.0\times 10^{-3}~{\rm eV}^2 $,
the $BR(\pi^0 \to \nu_L \bar \nu^i_R )$ is of the order $10^{-19}$ for 
$M_{\ast}\sim 1.4 $ TeV corresponding to $\l$ of $O(1)$.  
The situation remain unchanged even with the assumption of degenerate neutrino 
masses. In the scenario, where the right-handed neutrinos propagate in 
the extra-dimensions with largest size, the  
$BR(\pi^0 \to \nu_L \bar \nu^i_R )$ is too small to be observed.

\vskip .1in

{\large \bf B meson decays}

In the standard model, the process $ B \to \nu \bar \nu $ receives 
contributions from $Z$-penguin and box diagrams, where the dominant 
contribution comes from intermediate top quark loop. Off-shell $Z$ and 
$W$ exchanges have contributions from would be Goldstone modes that couple
to right handed neutrinos. The effective 
Lagrangian for $B \to \nu \bar \nu $ for each neutrino is given by
\beq
{\cal L}_{eff} = f_1 \Phi_B \nu \bar \nu 
\eqn
where, $f_1 = \frac{G_F}{\sqrt{2}}\frac{\alpha}{2\pi\sin^2\theta_W }
C^{SM}_{11} V^{*}_{td}V_{tb} F_B m$, $G_F$ is the Fermi coupling 
constants, $\alpha$ is the fine structure 
constant ( at the $Z$ mass scale), $F_B$ is the $B$-meson decay constant,
$\theta_W$ is the weak angle and $V^*_{td}V_{tb}$ are products of 
CKM matrix elements. The Wilson coefficient $C^{SM}_{11}$ at the leading 
order is given by:
\beq
C^{SM}_{11} = \frac{x_t}{8}\bigg[ \frac{x_t+2}{x_t-1}+ \frac{3(x_t -2)}{(x_t-1)^2}\ln(x_t) \bigg ]
\eqn
where $x_t = \frac{m^2_t}{m^2_W}$.

In models of large extra-dimension, $B \to \nu \bar \nu$ gets additional 
contribution from Higgs exchange contribution. The effective flavor changing 
vertex ($bdH^0$)can be obtained from \cite{Higgs_Hunter}
\beq
{\cal L}_{bdH^0} = \frac{G_F^{3/2}}{2^{1/4}}\frac{3}{16\pi^2}\sum_{i} m^2_i 
V^{*}_{id}V_{ib}\bigg [ m_b \bar d (1+\gamma_5) b + m_d \bar d (1+\gamma_5) b\bigg] H^0 + h.c.
\eqn
where, $V_{ij}$ are the elements of the Kobayashi-Maskawa matrix and $m_i$ 
are the corresponding quark masses flowing in the loop. 

The effective Lagrangian for the process $B \to \nu \bar \nu$ 
retaining only the top quark contribution is:             
%\beq
%{\cal L}_{eff} = \frac{G_F}{\sqrt{2}}\bigg[(\sqrt{2} G_F)^{1/2}
%\frac{F_B m_B^2}{m_b}\frac{m}{v}\frac{3 m_t^2 m_b}
%{16\pi^2 m^2_H}V^*_{td}V_{tb}
%\bigg ]\Phi_B \bar\nu^i_R \nu_L 
%\eqn
\beq
{\cal L}_{eff} = f_2 \Phi_B \bar\nu^i_R \nu_L 
\eqn
where, $f_2 = \frac{G_F} {\sqrt{2}}\bigg [ (\sqrt{2} G_F)^{1/2}
\frac{F_B m_B^2}{m_b}\frac{m }{v} V^{*}_{td} V_{tb} 
\frac{3 m_t^2 m_b}{16\pi^2 M^2_H} \bigg ] $
$M_H$ is the standard model Higgs mass and $v$ is the standard 
model vacuum expectation value. The decay width $B \to \nu_L \bar\nu^i_R$ 
(after summing over all KK modes of right handed neutrinos below $m_B$)
can be obtained after adding contributions from $Z$-penguin and box diagrams
and the Higgs mediated diagrams together
\beq
\Gamma (B \to \nu_L \bar\nu^i_R) = \frac{f^2_{total}}{16\pi } 
\bigg (\frac{m_B}{M_{*}}\bigg )^{\delta} \bigg (\frac{M_{Pl}}{M_{*}}\bigg )^2 
m_B
\eqn
where, $f_{total}= f_1 + f_2 $.
To compute the branching ratio we have taken following input 
parameters $m_b= 4.2$ GeV,
$m_B = 5.279 $ GeV, $m_t = 175 $ GeV, $v= 246$ GeV, $V_{td}= 0.006$,
$V_{tb}= 1.0$, $F_B= 0.180 $ GeV.
We then obtain the 
$BR( B \to \nu_L \bar \nu^i_R) $ for case of degenerate neutrinos with
$m_{\nu}=0.23 $ eV ( which also corresponds to the heaviest neutrino) and 
find that the branching ratio varies between 
$\sim 10^{-12}- 10^{-14}$ for $M_{\ast}\sim 50-120 $ TeV,
$\delta = 3 $ and $\l \sim 80- 20 $. This branching ratio is too small to be 
observed at any present $B$ factories. Even in the sub-space scenario, the
branching ratio does not get any significant enhancement, irrespective of 
the different neutrino masses, assumptions considered previously.
In the scenario, where the right-handed neutrinos propagate in
the extra-dimensions with largest size, the
$BR(B \to \nu_L \bar \nu^i_R )$ is too small to be observed.

%....................................................................%
%\section{Conclusions}
\vskip .1in      

{\large \bf Conclusions}

In this analysis we have studied the possible enhancement of invisible 
decay widths of the standard model Higgs boson and other pseudoscalar mesons
in the model of singlet neutrinos in extra-dimensions. In the case of
Higgs boson decay we have found that in certain range of extra-dimension
parameter space, the branching ratio of Higgs into invisible mode 
can be $\geq BR(H\to b \bar b) $. Unfortunately, the higher dimensional 
Yukawa coupling $\l$ takes on large $(\geq 100 )$ values in that parameter 
space. 
For $\l \leq 10 $, the $H\to \nu\bar\nu$
rate is a tiny fraction of the $H\to b\bar b$ rate. 

We have also studied the invisible 
decay rate of Higgs in the scenario, where right-handed neutrinos are in 
sub-space $( \delta_\nu < \delta )$, which is the modification of 
the minimal model, required to keep $\l \leq O(1) $. It has been shown that
to have a consistent model allowed by different experimental constraint
, one should
have $\delta_\nu = 5 $ and $\delta= 6$. In this scenario, the
invisible decay rate of Higgs can compete to that of $H \to b \bar b $, 
though for a very small range of $M_{\ast}\sim 1-1.3 $ TeV. 
In the scenario, where the right-handed neutrinos propagate in
the extra-dimensions with largest size, the invisible decay of Higgs 
can be as large as $H\to b \bar b$ for $M_H = 200 $ GeV, $\delta_\nu = 3,
\lambda = 1$ and $M_{\ast} = 1.5 $ TeV. However, for lower values of
$M_H$ ($= 100$ and $ 150 $ GeV) the invisible decay rate is smaller 
than $H\to b \bar b$ for the above set of parameters.

We have also studied the decay rates $\pi^0 \to \nu_L \bar \nu^i_R $ 
and $ B\to \nu_L \bar \nu^i_R$ in all these scenarios. Unfortunately in 
both of these decays the new effects are negligibly small to be measured.
\vskip .1in      

{\large \bf Acknowledgments}

This work was supported by US DOE contract numbers DE-FG03-96ER40969.
We would like to thank K. Agashe for reading the 
manuscript and making useful comments. 

%==================================================================%
\def\pr#1,#2 #3 { {Phys.~Rev.}        ~{\bf #1},  #2 (19#3) }
\def\prd#1,#2 #3{ { Phys.~Rev.}       ~{D \bf #1}, #2 (19#3) }
\def\pprd#1,#2 #3{ { Phys.~Rev.}      ~{D \bf #1}, #2 (20#3) }
\def\prl#1,#2 #3{ { Phys.~Rev.~Lett.}  ~{\bf #1},  #2 (19#3) }
\def\pprl#1,#2 #3{ {Phys. Rev. Lett.}   {\bf #1},  #2 (20#3)}
\def\plb#1,#2 #3{ { Phys.~Lett.}       ~{\bf B#1}, #2 (19#3) }
\def\pplb#1,#2 #3{ {Phys. Lett.}        {\bf B#1}, #2 (20#3)}
\def\npb#1,#2 #3{ {Nucl. Phys.}        {\bf B#1}, #2 (19#3)}
\def\pnpb#1,#2 #3{ {Nucl. Phys.}        {\bf B#1}, #2 (20#3)}
\def\prp#1,#2 #3{ { Phys.~Rep.}       ~{\bf #1},  #2 (19#3) }
\def\zpc#1,#2 #3{ { Z.~Phys.}          ~{\bf C#1}, #2 (19#3) }
\def\epj#1,#2 #3{ { Eur.~Phys.~J.}     ~{\bf C#1}, #2 (19#3) }
\def\mpl#1,#2 #3{ { Mod.~Phys.~Lett.}  ~{\bf A#1}, #2 (19#3) }
\def\ijmp#1,#2 #3{{ Int.~J.~Mod.~Phys.}~{\bf A#1}, #2 (19#3) }
\def\ptp#1,#2 #3{ { Prog.~Theor.~Phys.}~{\bf #1},  #2 (19#3) }
\def\jhep#1, #2 #3{ {J. High Energy Phys.} {\bf #1}, #2 (19#3)}
\def\pjhep#1, #2 #3{ {J. High Energy Phys.} {\bf #1}, #2 (20#3)}
%....................................................................%

%************************************************************************%
\begin{figure}[hbt]
\centerline{\epsfig{file=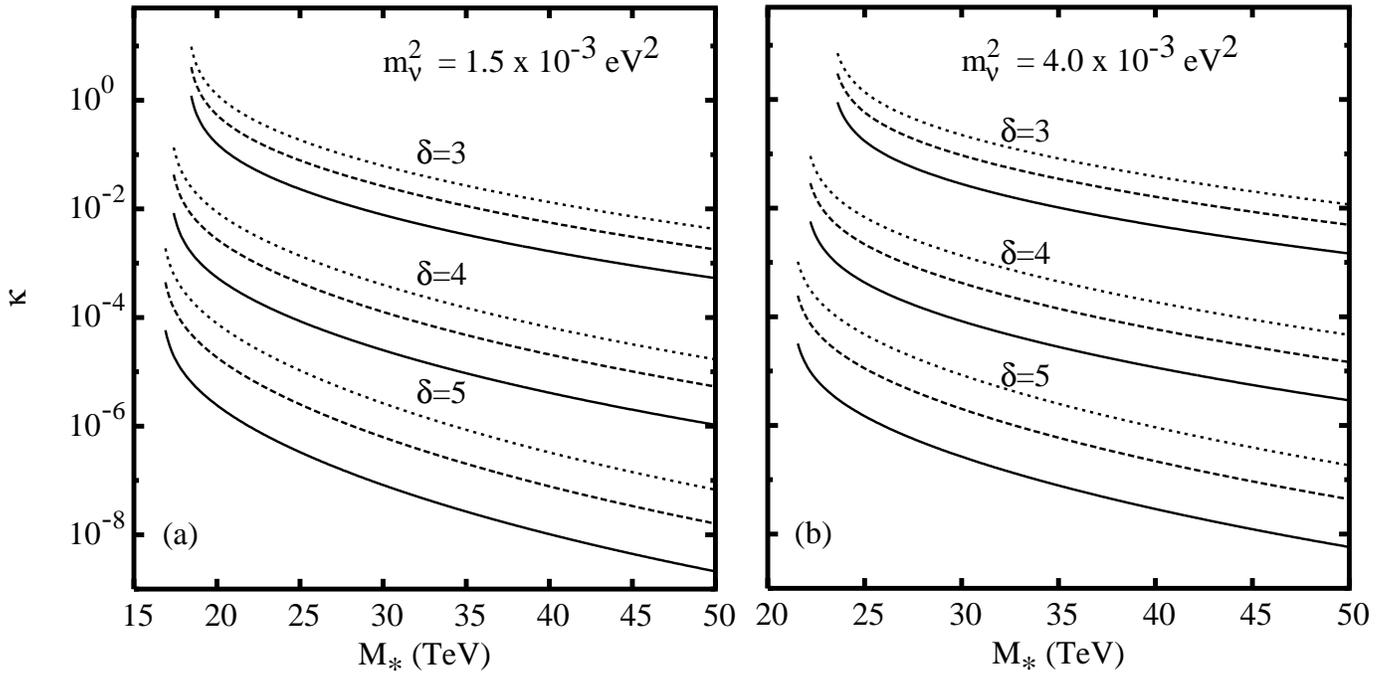,width=\linewidth}}
\vspace*{-8.0cm}
\caption{ Variation of $\kappa$ (defined in the text) with $M_{*}$ for 
three values of extra dimensions $\delta =3,4$ and 5. The solid, broken
and dotted lines correspond to $M_H = 100, 150 $ and 200 GeV respectively.
Figures $(a)$ and $(b)$ correspond to $m^2_{\nu}= 1.5\times 10^{-3}{\rm eV^2}$
and $ 4.0\times 10^{-3}{\rm eV^2}$ respectively.}
\label{fig1}
\end{figure}
\begin{figure}[hbt]
\centerline{\epsfig{file=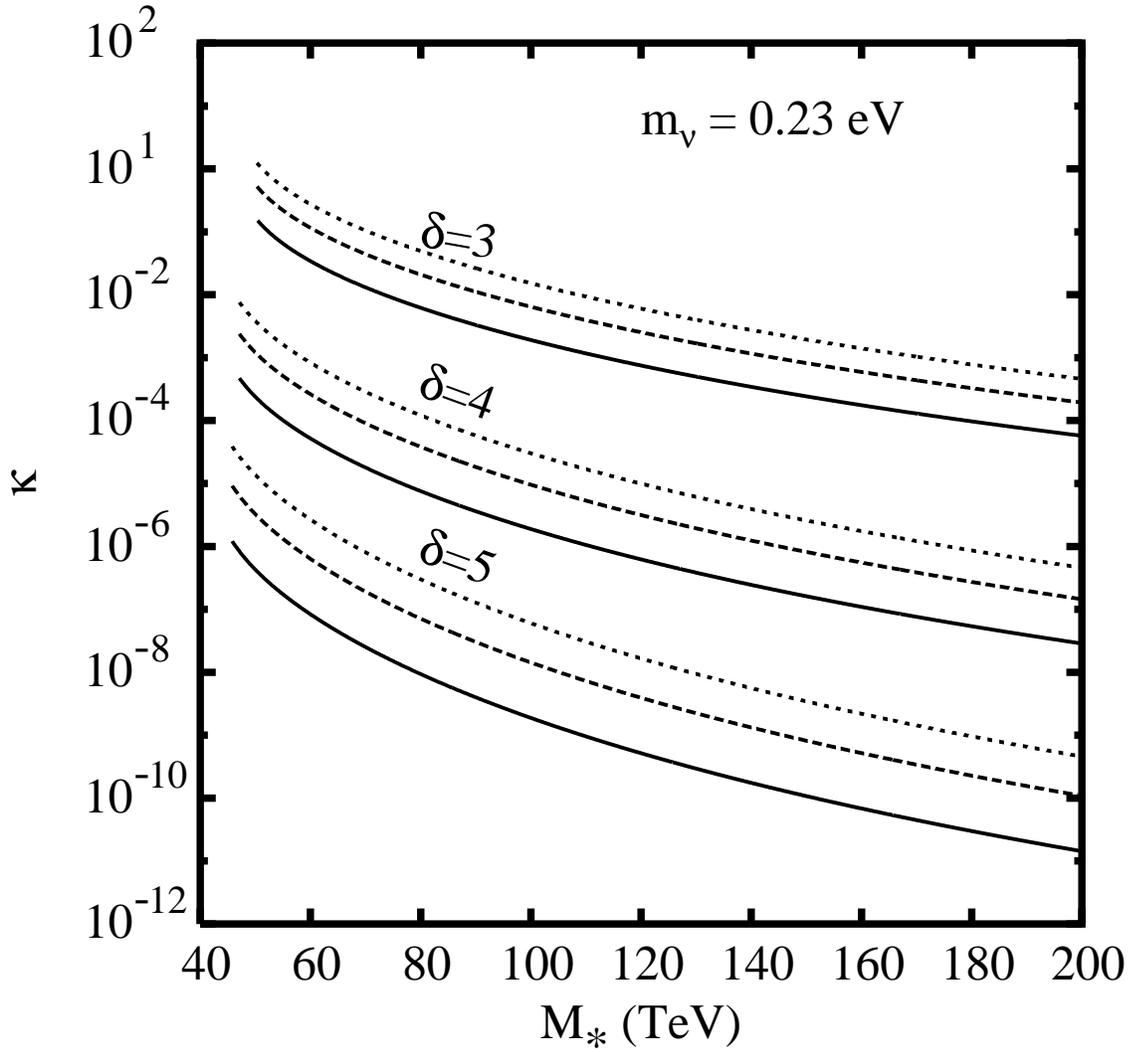,width=\linewidth}}
\vspace*{-8.0cm}
\caption{ Variation of $\kappa$ with $M_{*}$ for 
three values of extra dimensions $\delta =3,4$ and 5. The solid, broken
and dotted lines correspond to $M_H = 100, 150 $ and 200 GeV respectively.
We have fixed $m_{\nu}= 0.23 $ eV.}
\end{figure}

\begin{figure}[hbt]
\centerline{\epsfig{file=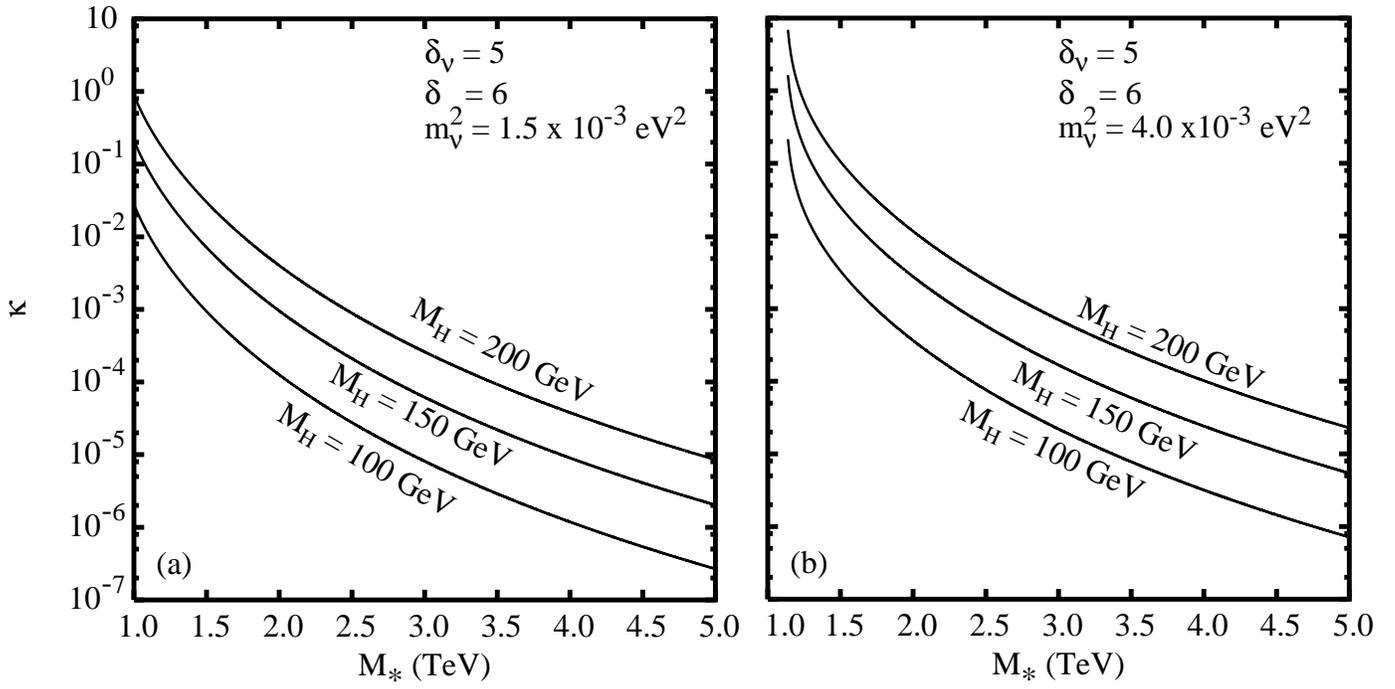,width=\linewidth}}
\vspace*{-8.0cm}
\caption{ Variation of $\kappa$ with $M_{*}$ for 
number of sub-space extra dimensions $\delta_{\nu} =5$,  
number of extra-dimensions $\delta= 6$ and $M_H = 100 $, 150 and 200 GeV.
Figures $(a)$ and $(b)$ correspond to 
$m^2_{\nu}= 1.5\times 10^{-3}{\rm eV^2}$
and $ 4.0\times 10^{-3}{\rm eV^2}$ respectively.}
\end{figure}
\begin{figure}[hbt]
\centerline{\epsfig{file=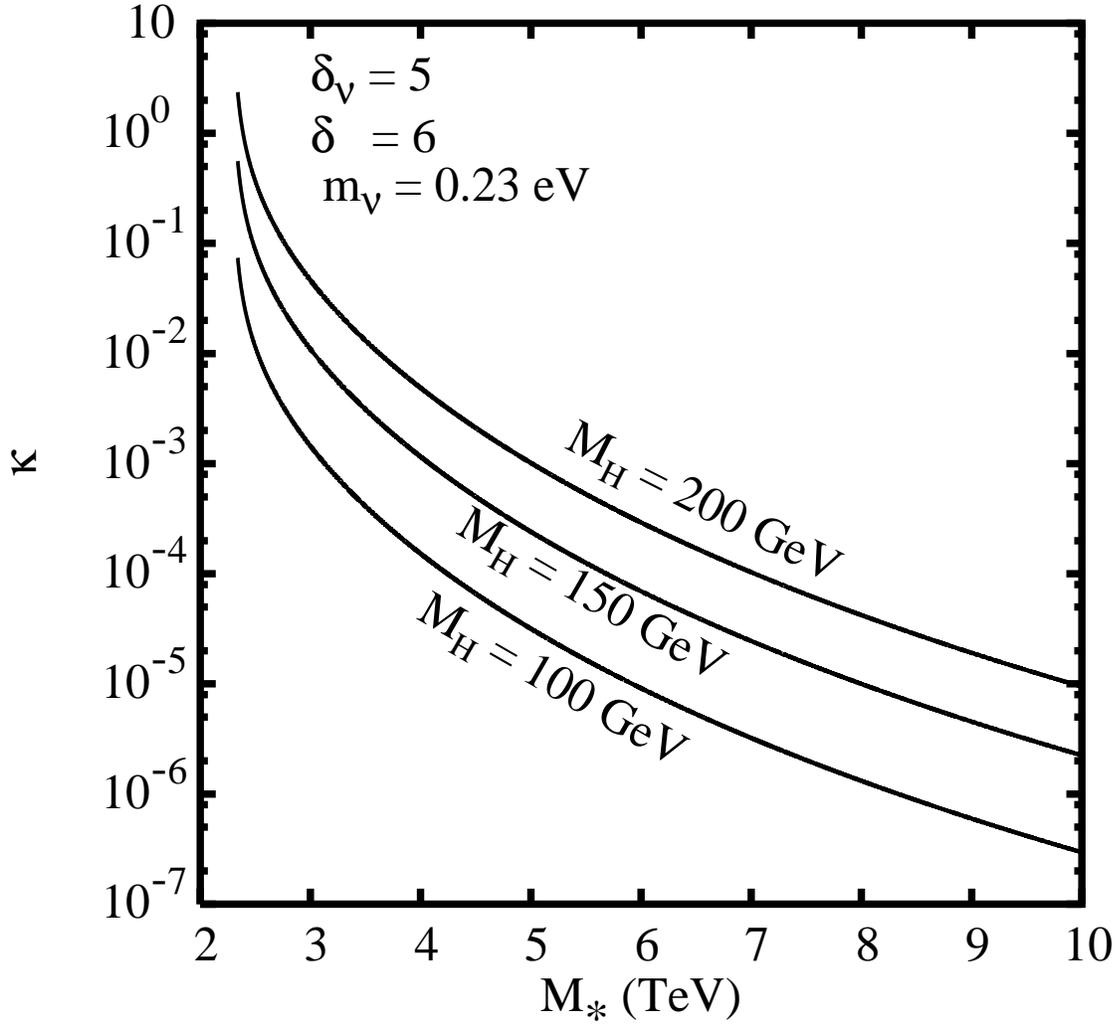,width=\linewidth}}
\vspace*{-8.0cm}
\caption{Variation of $\kappa $ as a function of 
$M_{\ast}$ in sub-space extra-dimensions $\delta_\nu= 5$,  
number of extra-dimensions $\delta = 6 $ and $M_H= 100$, 150 and 200 GeV. 
We have fixed $m_{\nu}= 0.23 $ eV.}
\end{figure}
\end{document}